\documentclass[journal]{IEEEtran}

\usepackage{amsmath,amssymb,amsfonts}
\usepackage{algorithmic}
\usepackage{graphicx}
\usepackage{textcomp}
\usepackage{braket}

\usepackage{soul,color}
\usepackage{algorithm}
\usepackage{array}
\usepackage{url}
\usepackage{multirow}
\usepackage{float}
\usepackage{hyperref}
\newcommand{\Mod}[1]{\ (\mathrm{mod}\ #1)}

\def\BibTeX{{\rm B\kern-.05em{\sc i\kern-.025em b}\kern-.08em
    T\kern-.1667em\lower.7ex\hbox{E}\kern-.125emX}}
\usepackage[
backend=biber,
style=ieee,
citestyle=numeric-comp,
]{biblatex}    
\begin{document}
\title{Reducing the Depth of Quantum FLT-Based Inversion Circuit}
\author{Harashta Tatimma Larasati, Dedy Septono Catur Putranto, Rini Wisnu Wardhani, Howon Kim\\
\{harashta, dedy.septono, rini.wisnu, howonkim\}@pusan.ac.kr

}

\maketitle

\begin{abstract}
Works on quantum computing and cryptanalysis has increased significantly in the past few years. Various constructions of quantum arithmetic circuits, as one of the essential components in the field, has also been proposed. However, there has only been a few studies on finite field inversion despite its essential use in realizing quantum algorithms, such as in Shor's algorithm for Elliptic Curve Discrete Logarith Problem (ECDLP). In this study, we propose to reduce the depth of the existing quantum Fermat's Little Theorem (FLT)-based inversion circuit for binary finite field. In particular, we propose follow a complete waterfall approach to translate the Itoh-Tsujii's variant of FLT to the corresponding quantum circuit and remove the inverse squaring operations employed in the previous work by Banegas et al., lowering the number of CNOT gates (CNOT count), which contributes to reduced overall depth and gate count. Furthermore, compare the cost by firstly constructing our method and previous work's in Qiskit quantum computer simulator and perform the resource analysis. Our approach can serve as an alternative for a time-efficient implementation.
\end{abstract}

\begin{IEEEkeywords}
quantum cryptanalysis, inversion, binary field, simulation.
\end{IEEEkeywords}

\IEEEpeerreviewmaketitle

\section{Introduction}
\label{sec:intro}
The study on quantum computing has been emerging, particularly after Peter Shor demonstrated the apparent advantage of using quantum phenomena to speed up computation and crack classically intractable problems underlying current public-key cryptosystems, later known as the Shor's algorithm \cite{shor1994algorithms,shor1999polynomial}. In addition, research in quantum hardware has seen significant advancement in the past few years, continually achieving an increased number of physical qubits every few years \cite{cho_2020, ball_2021}. This accelerates the potential use of quantum computers in Noisy Intermediate-Scale Quantum (NISQ) era, and later in the Fault Tolerant Quantum Computation (FTQC). Inevitably, the construction of efficient quantum arithmetic circuits, i.e., circuits to perform arithmetic operations in a quantum computer, becomes essential. 

To date, numerous research efforts have proposed the quantum circuit implementations of arithmetic operations. Driven by the interest in Shor's algorithm for factoring \textemdash the variant to crack Rivest–Shamir-Adleman (RSA)-based cryptosystems, early research \cite{vedral1996quantum,beckman1996efficient} had focused on designing an explicit construction of quantum circuits for performing \textit{modular exponentiation}, one of the essential components comprised in the algorithm. The circuits include  specialized circuits and their underlying components, such as adder, modular adder, and modular multiplier by a constant. They are further improved in the later research customized to certain properties of modular arithmetic operations to obtain additional optimizations, e.g., \cite{beauregard2002circuit, markov2012constant,van2005fast,rines2018high}.

Apart from for factoring, the other variant of Shor's algorithm as the basis for cracking Elliptic Curve Cryptography (ECC)-based cryptosystem has also been studied, calling for concrete construction of its underlying circuit as well, e.g., \cite{proos2003shor,roetteler2017quantum,haner2020improved,banegas2021concrete}, though not as flourished. For this case, the circuit building blocks pose differences from the previous algorithm variant. Instead of a \textit{modular exponentiation}, Shor's algorithm for Elliptic Curve Discrete Logarithm Problem (ECDLP) requires a \textit{scalar multiplication}, which is further can be broken down into a series of \textit{point addition} circuit. One of the essential component of the point addition circuit is the \textit{inversion} block, used together with multiplication block to perform a \textit{division} in the finite field case. In terms of the type of finite field and its respective elliptic curve, the works in quantum cryptanalysis of Shor's ECDLP can be classified into two main areas: for prime elliptic curves, such as \cite{proos2003shor,roetteler2017quantum,haner2020improved} and for binary elliptic curves \cite{rotteler2013quantum,budhathoki2015automatic,banegas2021concrete}.

In terms of inversion operation, there has only been a few methods proposed for the quantum case, some of which are based on the classical extended Euclidean algorithm (EEA) \cite{proos2003shor} and its variant \cite{roetteler2017quantum, haner2020improved}, extended greatest common divisor (GCD) \cite{banegas2021concrete}, and Fermat’s Little Theorem (FLT) \cite{banegas2021concrete}. As for the binary elliptic curves, the latter two are the most recent with an explicit quantum circuit construction.
Compared to other arithmetic operations such as addition and multiplication, the amount of proposals to improve inversion circuit is very minimal despite its significance in the quantum cryptanalysis and quantum arithmetic in general. To meet different needs, e.g., space-efficient implementation, time-efficient implementation, or combination of both), an alternative to the existing approach will be very beneficial.

In this paper, we propose to reduce the depth of the FLT-based quantum inversion circuit for binary elliptic curves. We depart from the previous work by Banegas et al. \cite{banegas2021concrete}, which interprets the second variant of classical FLT-based Itoh-Tsujii algorithm to the corresponding quantum circuit, adjusting their algorithm for a more time-efficient implementation. In particular, we employ a complete waterfall approach and alter the first and second stage of their algorithm, relocating the uncomputation to the end, thus eliminating the intermediate uncomputation (i.e., repeated inverse squarings) used in the previous work and minimizing the number of CNOT gates in the circuit. As a result, an overall lower depth and gate count can be achieved. Additionally, for verification and comparison, we build the code for both our proposed inversion method and the previous work in Qiskit quantum computer simulator, then perform the resource analysis for the inversion bit size $n$ of 8, 16, 127, 163, and for the use case of binary elliptic curve standard: B-233, B-283, B-409, and B-571 as, for the context of quantum cryptanalysis. Note that our proposed method comes with a tradeoff of higher ancilla qubits. Nevertheless, by the fact that the depth corresponds to the time complexity of a quantum circuit \cite{gyongyosi2020circuit} and considering the potential benefit later discussed in Section \ref{sec:discussions}, this work is advantageous for a time-efficient implementation.

\section{Preliminaries}
\label{sec:prelim}
\subsection{Binary Finite Fields in Classical \& Quantum Computing}
Finite fields, or fields with a finite number of elements, 
are commonly employed in cryptographic applications such as symmetric and asymmetric key cryptosystems and have applications in various other domains such as network coding and error control theory \cite{NadikudaClassicalInversion}. The binary field GF($2^m$) \textemdash here we refer to as GF($2^n$) \textemdash and prime field GF($p$) can be considered as the most extensively utilized finite fields. For a detailed comparison of prime and binary fields, readers can refer to \cite{wenger2011exploring}.

Regarding binary fields, for classical computer implementation, a simple logical exclusive-OR (XOR) gate is used to perform addition, whereas a logical AND gate is often used to conduct multiplication.
The binary fields are more easily employed than prime fields due to their carry-free characteristic and more staightforward hardware implementation. 

Translating to its quantum counterpart, performing quantum arithmetic operation in binary field is also relatively simpler and more cost-efficient than in prime field since addition can be performed by just a CNOT gate for each bit, whereas multiplication can be done directly by a series of Toffoli gates (plus reduction by a series of swap gates and small amount of CNOT gates). 

In a quantum computer environment, the depth of a quantum circuit refers to the number of time steps (time complexity) required for the quantum operations making up the circuit to run \cite{gyongyosi2020circuit}, which is important since maintaining a long coherence in a quantum computer is still a very challenging problem.

\subsection{Related Works on Quantum Inversion Circuit}
Despite the importance of inversion circuit in the quantum cryptanalysis, there has not been many works that extends beyond the specific implementation and optimization for small bits, such as for Advanced Encryption Standard (AES) inversion which is fixed on 8-bit inversion, e.g., \cite{etri2020towards}. 
In terms of a more general technique or for larger numbers, such as for use in Shor's ECDLP for ECC quantum cryptanalysis, there has only been a few to date. 

For prime fields (i.e., $GF(p)$), the pioneering work is in 2003 by Proos and Zalka \cite{proos2003shor}, which started the discussion on using extended Euclidean algorithm (EEA) for performing the inversion. Fourteen years later, quantum computer simulators began to rise in development, enabling a more fine-grained, gate-level optimization of quantum circuit. That time, Roetteler et al. \cite{roetteler2017quantum} proposed another approach for performing inversion, i.e,, using the Kaliski's almost inverse (also known as the Kaliski's binary GCD) algorithm, since they found out that constructing an efficient EEA-based quantum inversion circuit is far from trivial.  

Regarding the work in the binary elliptic curves, early research in the past decade (e.g., \cite{cheung2008design,maslov2009m2,rotteler2013quantum}) focused on employing the projective coordinate to eliminate the need of inversion circuit. However, as pointed out in the more recent literature, the division/inversion problem can not be completely removed since unlike in the affine coordinates, projective coordinates introduce new challenge of \textit{non-unique} representation of points, which is required in Shor's algorithm \cite{haner2020improved}. To re-obtain a unique representation, one will still need to perform division by \textit{Z} coordinate, which is also expensive \cite{haner2020improved}.

In terms of the more recent works for quantum inversion in binary elliptic curves, two different approach are recently proposed by Haner et al, \cite{haner2020improved}: extended GCD algorithm variant adopted from the classical inversion by \cite{bernstein2019fast}, and a Fermat's Little Theorem (FLT)-based inversion adapted from classical Itoh-Tsujii inversion \cite{itoh1989structure}. Readers interested in a more detailed research landscape of the works in Shor's ECDLP may refer to \cite{larasati2021quantum}.

\subsection{FLT-based Inversion in Quantum Computing}
\label{subsec:flt}
The use of Fermat's Little Theorem (FLT)-based inversion for quantum computing and cryptanalysis has first been discussed by Banegas et al \cite{banegas2021concrete}. The theorem itself, introduced by Pierre de Fermat in 1640, states that for a prime number $p$ and any number not divisible by $p$, say $f$, their relation can be described as Equation \ref{eq:flt_basic_1}. 
\begin{equation}\label{eq:flt_basic_1}
    f^{p-1}\equiv1 \Mod{p}
\end{equation}
If both sides are multiplied with $a^{-1}$, the result is as stated in Equation \ref{eq:flt_basic_2}. 
\begin{equation}\label{eq:flt_basic_2}
    f^{-1}\equiv f^{p-2} \Mod{p}
\end{equation}
That is, an inversion $f^{-1}$ can be performed by utilizing exponentiation. This also applies to binary finite fields, in which the equation can be slightly rewritten as in Equation \ref{eq:flt_basic_binary}.
\begin{equation}\label{eq:flt_basic_binary}
    f^{-1}\equiv f^{2^n-2} \Mod{m(x)}
\end{equation}
 In particular, the inversion can be achieved using $n$ multiplications and $n-1$ squarings \cite{banegas2021concrete}, as shown in Equation \ref{eq:flt_basic_binary_long}.

\begin{equation}\label{eq:flt_basic_binary_long}
f^{2^{n}-2}=f^{2^{1}}.f^{2^{2}}.f^{2^{3}}...f^{2^{n-1}}.  
\end{equation}

Classically, inversion using the Fermat's Little Theorem (FLT) approach is generally less preferable than the extended Euclidean counterpart because to perform exponentiation, a large number of multiplications need to be employed, hence considered more expensive. Nevertheless, several papers have shown that using the right combination and approach, FLT-based inversion can also contribute to higher speed, such as in \cite{awaludin2021high}. Furthermore, there exists improved variants of FLT algorithm, with the most popular are the ones proposed by Itoh and Tsujii \cite{itoh1989structure}, which reduces the number of multiplications via a smart arrangement of equations. 


For the use in the quantum binary finite fields, deriving from classical FLT-based inversion algorithm, Banegas et al. \cite{banegas2021concrete} has described the construction of the corresponding quantum circuit. In particular, they follow the well-known Itoh-Tsujii’s derivation of the FLT algorithm which requires a smaller number of operations than the standard derivation. Using the second variant of Itoh-Tsujii, the inversion cost can be reduced to below $2\log(n)$ multiplications and with the same amount of squarings as the original FLT \cite{banegas2021concrete}. In detail, lowering the amount of multiplications can be achieved by considering the following two distributive equations: 
\begin{equation} \label{eq:distrib1} 
    f^{2^{n}-2} = {(f^{2^{n-1}-1})}^2
\end{equation}
\begin{equation} \label{eq:distrib2}
    (f^{2^{2^t}-1} = {f^{2^{2^t}-1})}^{2^{2^{t-1}}} (f^{2^{2^t}-1})
\end{equation}
    
Let $n-1$ be written as $k_1\dots k_t$ with $\sum_{s=1}^{t}{2^{k_s}}= n-1$ and $k_1>k_2>k_3> \dots k_t \geq 0$, with $t$ the Hamming weight of $n-1$ in binary, $t\leq\lfloor \log_{}{(n-1)}\rfloor+1$, and $k_1 = \lfloor \log_{}{(n-1)}\rfloor$. Using Equations \ref{eq:distrib1} and \ref{eq:distrib2}, inversion via exponentiation can be achieved through the following three stages \cite{banegas2021concrete}: 
\begin{enumerate}
\item Calculate $f^{2^{2^{k_1}-1}}$ with $k_1$ multiplications using Equation \ref{eq:distrib2}, save the intermediate result $f^{2^{2^{k_t}-1}}$, $f^{2^{2^{k_{t-1}}-1}}$, \dots, $f^{2^{2^{k_1}-1}}$. 
\item Calculate 
$
\big
\{
\dots\big\{(f^{2^{2^{k_1}-1}})^{2^{2^{k_2}}} (f^{2^{2^{k_2}-1}})\}^{2^{2^{k_3}}}\dots\}^{2^{2^{k_t}}}\\(f^{2^{2^{k_t}-1}})
\}$
using $t-1$ multiplications.
\item Square 1) and 2) to obtain the inverse, $f^{-1}$.
\end{enumerate}





In their paper, Banegas et al. \cite{banegas2021concrete} describe the quantum cryptanalysis of binary elliptic curve. Specifically, the describe the implementation of Shor's ECDLP, focusing on the point addition operations and particularly the underlying components, including the division operation which can be done by an inversion followed by a multiplication in the respective registers. Their target of optimization is the circuit width (i.e., qubit size). For their FLT-based inversion, their translation of Itoh-Tsujii's algorithm to the quantum circuit are presented at Algorithm 2 in the Section 6.2 of their paper \cite{banegas2021concrete}. In summary, to perform the inversion, they employ a series of squarings (i.e, $K$ in their paper), (modular) multiplications $M$, along with the inverse squarings $K^{-1}$. The example of their construction is illustrated on Circuit 6 in the Section 6.2 of their paper.   
In addition, there has been several works that focuses on reducing depth rather than width, such as by Rahman et al., \cite{rahman2022grover}, which constructs a combinatorial circuit to analyze KATAN block cipher by optimizing the depth and number of gates \cite{rahman2022groverkatan}, and depth analysis in \cite{reversible_de2021reducing}. 

\section{Proposed Method} 
\label{sec:proposed_method}
In this section, we describe our proposed quantum FLT-based inversion operation for binary elliptic curve $GF(2^n)$ to achieve a lower depth. In particular, we propose a different approach from the previous work by Banegas et al. \cite{banegas2021concrete}; that is, here we employ a complete waterfall approach to translate the Itoh-Tsujii algorithm to the corresponding quantum circuit. We modify the steps in previous work \cite{banegas2021concrete} to minimize in the number of CNOT gates. As a result, the overall depth and gate count can be lowered whereas the same T-depth  as the previous work can be maintained.

\subsection{Proposed Variant of FLT-based Inversion}
In this paper, to aid the explanation of our method, we start by elaborating the quantum circuit construction used in Banegas et al \cite{banegas2021concrete}, then introduce our differences along the way.

\begin{figure}[h] 
    \centering
    \includegraphics[width=1\linewidth]{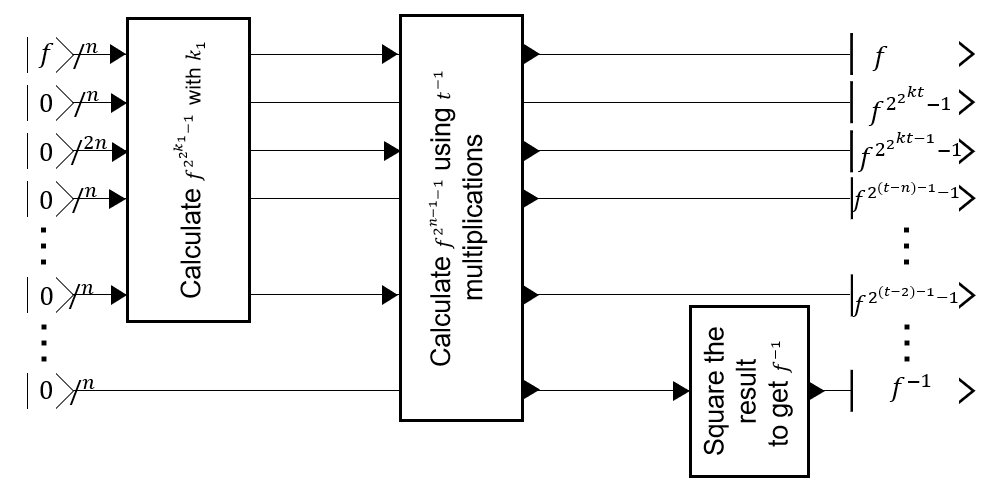}
  \caption{A three-stage FLT inversion steps, inferred from \cite{banegas2021concrete}}
  \label{fig:flt_3stage}
\end{figure}

In general, the quantum FLT-based inversion consists of three stages \cite{banegas2021concrete}, as discussed in the Section \ref{subsec:flt}
and illustrated in Fig. \ref{fig:flt_3stage}. At the first stage, $f^{2^{2^{k_1}}-1}$ is calculated. This can be performed by a series of squarings and (modular) multiplications. At the second stage, $f^{2^{(n-1)}-1)}$ is calculated in a similar fashion. At the third stage, both are squared to obtain the inversion result, $f^{-1}$. 

In the method implemented in the previous work \cite{banegas2021concrete}, an inverse squaring operation (namely, $K^{-1}$, or equivalently $SQUARE^{-1}$) is required in each iteration of the first stage. That is, after finishing a squaring, the value is uncomputed by the $SQUARE^{-1}$ so that the qubit can be reused for the subsequent squaring. Unfortunately, this introduces more CNOT gates to the circuit since $SQUARE^{-1}$ practically consists of CNOT gates, which is bounded by $O(n^2)$. Furthermore, qubit reuse also requires additional series of CNOT gates of at least $n$ per squaring for uncomputation, yielding larger overall depth and gate count.

In this paper, a more straightforward way of using a waterfall approach to perform inversion is presented. We keep the standard method of using three stages of calculation. However, the method of calculation is different, giving different construction and operation placement, as presented in Algorithm \ref{algo:flt}. Furthermore, using a sequential arrangement of values, uncomputation is performed after the calculation finishes; hence $SQUARE^{-1}$ and additional CNOTs can be removed. As a result, a circuit with a lower overall depth and smaller gate count can be obtained. 

Regarding the explanation of Algorithm \ref{algo:flt}, the algorithm is for performing a time-efficient FLT-based quantum inversion circuit for a binary elliptic curve using constant field polynomial $m(x)$ of degree $n>0$ as a fixed input. For simplicity and ease of comparison, we closely follow the notation and description of \cite{banegas2021concrete}. $k_{1}$ is a register that saves result multiplication final in stage 1 as a predecessor value for the next stage, while $s$ is the iteration variable. Note that $k_1>k_2> \dots k_t\geq0$ such that $\sum_{s=1}^{t}{2^{k_s}}= n-1$. In this case, maximum register number is $k=2k_1+t$, in quantum circuit we define this value as number of qubits.

For the operations (functions), lower case represents the standard gate operation (in this case, CNOT and swap), while the all-capital case (in this case, \textit{SQUARE} and \textit{MULT}) represents a block of operation which can be rolled out according to one’s implementation preference. Generally, for binary finite field quantum circuit, squaring (\textit{SQUARE}) operation is constructed via LUP Decomposition, a decomposition method for linear mapping also used in \cite{etri2020towards, banegas2021concrete}. As for the (modular) multiplication operation, the straightforward Schoolbook multiplication (followed by reduction) or other approach such as Karatsuba or Toom-Cook can also be employed. In this study, since our focus is on the inversion algorithm itself, we utilize the simple-yet-straightforward Schoolbook multiplication for the underlying multiplication block.

\begin{algorithm} 
\caption{Our proposed variant of FLT. For simplicity and ease of comparison, notation and description follow \cite{banegas2021concrete}.}
\begin{algorithmic}\label{algo:flt}
\raggedright
\STATE {\textbf{Fixed input} : A constant field polynomial $m(x)$ of\\}
\setlength{\leftskip}{2.6cm}
\STATE {degree $n>0. k_1 > k_2 > \dots k_t\geq 0$ such that$\sum\limits_{s=1}^3{2^{k_s}=n-1}. k_{max}=2*k_1 + t.$}\\
\setlength{\leftskip}{0cm}
\setlength{\rightskip}{0cm}
\STATE {\textbf{Quantum input} : \\}
\setlength{\leftskip}{0cm}
\STATE {- A non-zero binary polynomials of degree up to \\}
\setlength{\leftskip}{0.3cm}
\STATE {$n - 1$ stored in array (register) $f_0$ of size $n$ to invert.\\}
\setlength{\leftskip}{0cm}
\setlength{\leftskip}{0.3cm}
\setlength{\leftskip}{0cm}
\STATE {- $k$ zero arrays of size $n$ initialized to an all-$\ket{0}$ \\}
\setlength{\leftskip}{0.3cm}
\STATE {state: $f_1, \dots , f_k$.\\}
\setlength{\leftskip}{0cm}
\setlength{\rightskip}{0cm}
\STATE {\textbf{Result} : inverse of the input, stored in $f_k$\\}
\STATE {1: } \textbf{for} $i=1 ,… ,k_1$  \textbf{do}\hfill//stage 1
\STATE {2: } \hspace{5mm} CNOT$(f_{2*(i-1)+1},f_{2*(i-1)})$ 
\STATE {3: } \hspace{5mm} \textbf{for} $i=1,… , t-1$  \textbf{do}
\STATE {4: } \hspace{10mm} SQUARE$(f_{2*(i-1)+1})$ 
\STATE {5: } \hspace{10mm} MULT$(f_{2*(i-1)+2},f_{2*(i-1)+1,f_{2*(i-1)})}$ 
\STATE {6: } \textbf{for} $s=1 ,… ,t-1$  \textbf{do} \hfill//stage2
\STATE {8: } \hspace{5mm} \textbf{for} $k=1 ,… ,2^{k_{s+1}}$  \textbf{do}
\STATE {9: } \hspace{10mm} SQUARE$(f_{2*(i-1)+1})$ 
\STATE {10: } \hspace{8mm} MULT$(f_{2*(i-1)+2},f_{2*(i-1)+1,f_{2*(i-1)})}$ 
\STATE {11: } \textbf{if} $t=1$ \textbf{then}  
\STATE {12: } \hspace{5mm} swap $(f_{k_1}, f_k)$
\STATE {13: } SQUARE$(f_k)$ \hfill//stage 3
\end{algorithmic}
\end{algorithm}

Note that more ancilla qubits are required compared to the previous work, but they similarly are to be uncomputed at the end. In addition, considering the development of quantum hardware which has seen a rapid acceleration in the increasing number of qubits, this approach does pose a competitive advantage.

\begin{figure} [h]
    \centering
    \includegraphics[width=1\linewidth]{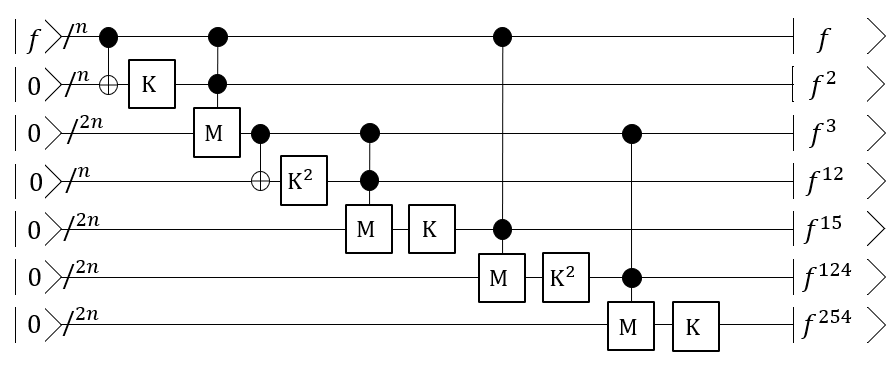}
  \caption{Illustration of Modified FLT Algorithm, Example for $n = 8$}
  \label{fig:flt_scheme_8bit}
\end{figure}


\subsection{Experiment Setup for Evaluation}
To compare the performance of our method and previous work, we construct both the FLT inversion by Banegas et al., \cite{banegas2021concrete} and ours in the Qiskit quantum computer simulator platform and run the resource analysis. For the previous work's construction, we follow the algorithm presented on Algorithm 2 in their paper.
In our experiment, since our scope is on the FLT inversion evaluation and not the lower-level component (e.g., the multiplication block), we employ the exact same setting for the  underlying circuits (e.g., same version of squaring and multiplication blocks) for both Banegas et al's and ours. Additionally, for constructing a modular multiplication, we utilize the Schoolbook multiplication followed by reduction for both scenarios due to its simplicity. Employing other multiplication methods (e.g., quantum Karatsuba such as \cite{banegas2021concrete,parent} or Toom Cook multiplication such as \cite{dutta,larasati2021toom}) can be done accordingly by simply changing the underlying blocks since it is supported by Qiskit. Following previous works on quantum cryptanalysis \cite{rahman2022groverkatan, haner2020improved, banegas2021concrete}, swap operations can be considered as free since it can be done via qubit relabeling \cite{jang2020grover, jang2021efficient}.

Furthermore, since the most prominent use of this quantum inversion circuit is on the quantum cryptanalysis of binary elliptic curve using Shor's ECDLP, we evaluate our work with the parameters for several NIST's standardized curves, i.e., B-233, B-283 B-409 and B-571 with their respective irreducible polynomials, along with the smaller curve with the irreducible polynomial listed in \cite{banegas2021concrete}. In terms of the metric used, we currently analyze the number of CNOT Gates (i.e., CNOT count), circuit depth, and qubit size (i.e., circuit width).

\section{Result and Discussions}
\label{sec:result_discussion}
\subsection{Evaluation Result}
To verify our proposed method, we implement both versions of FLT-based algorithm (i.e., ours and Banegas et al.,'s \cite{banegas2021concrete}) on Qiskit platform. Then, we perform the resource analysis on the decomposed (i.e., transpile function in Qiskit) version of the quantum circuits and compare the result. We ran our experiment on Python 3.9.7, Qiskit version 0.30.0 locally in an x64-based desktop PC (Intel i7-8700, 6 cores processor), on Windows 10 Pro OS with 64 GB of RAM. For the transpile function, the basis gates follow the decomposition described in Q-Crypton, a quantum-security evaluation platform for cryptography developed by ETRI, a Korean government research institute. In particular, the allowed basis gates are ['id', 'h','t', 'tdg', 's', 'sdg', 'rz', 'x', 'y', 'z', 'cx'], without utilizing optimization function (i.e., set the optimization level to 0). 

\begin{figure}[h]
    \centering
    \includegraphics[width=\linewidth]{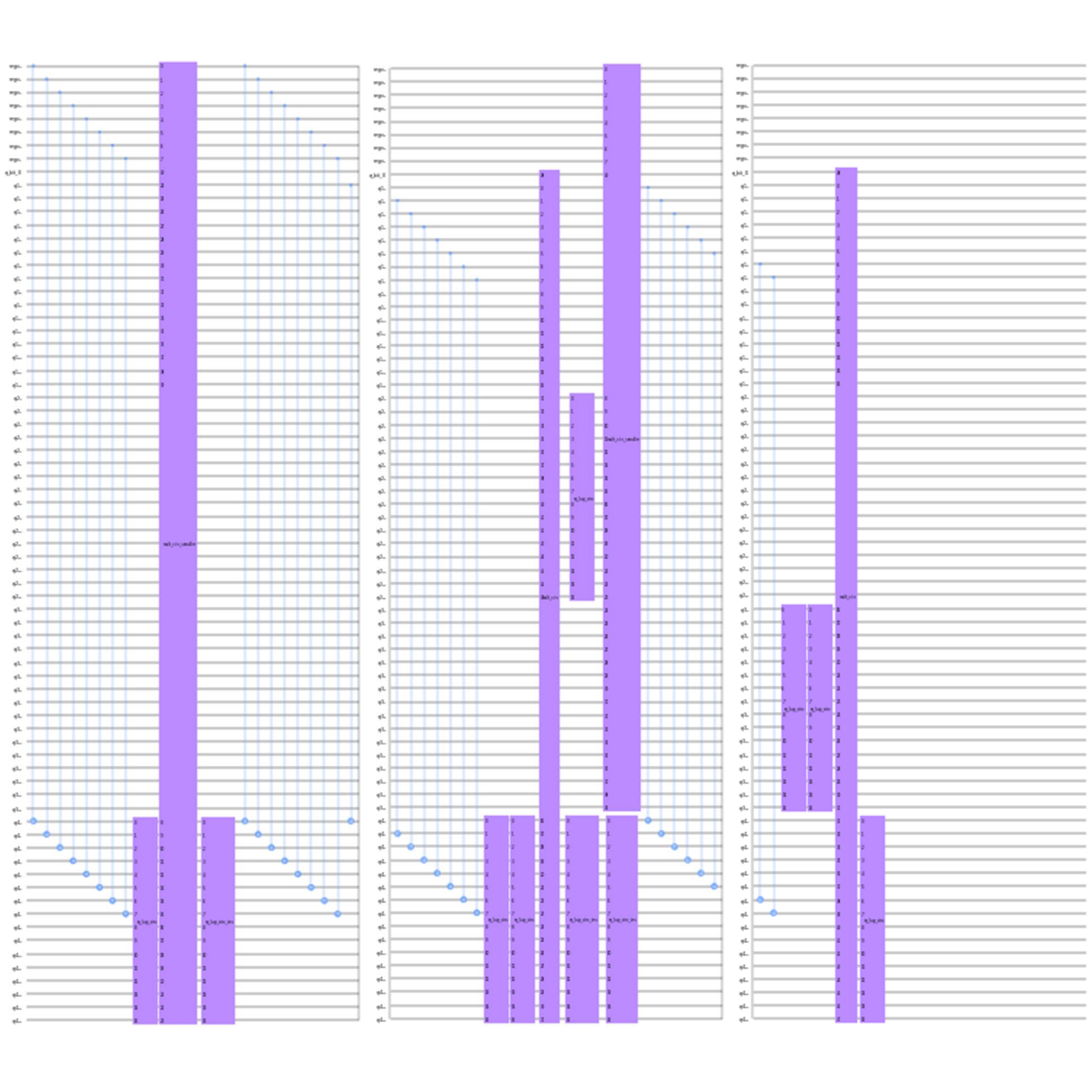}
  \caption{Our Qiskit's Circuit Construction of Previous Work's FLT-based Inversion \cite{banegas2021concrete}}
  \label{fig:qiskit_prevwork}
\end{figure}
\begin{figure}[h]
    \centering
    \includegraphics[width=\linewidth]{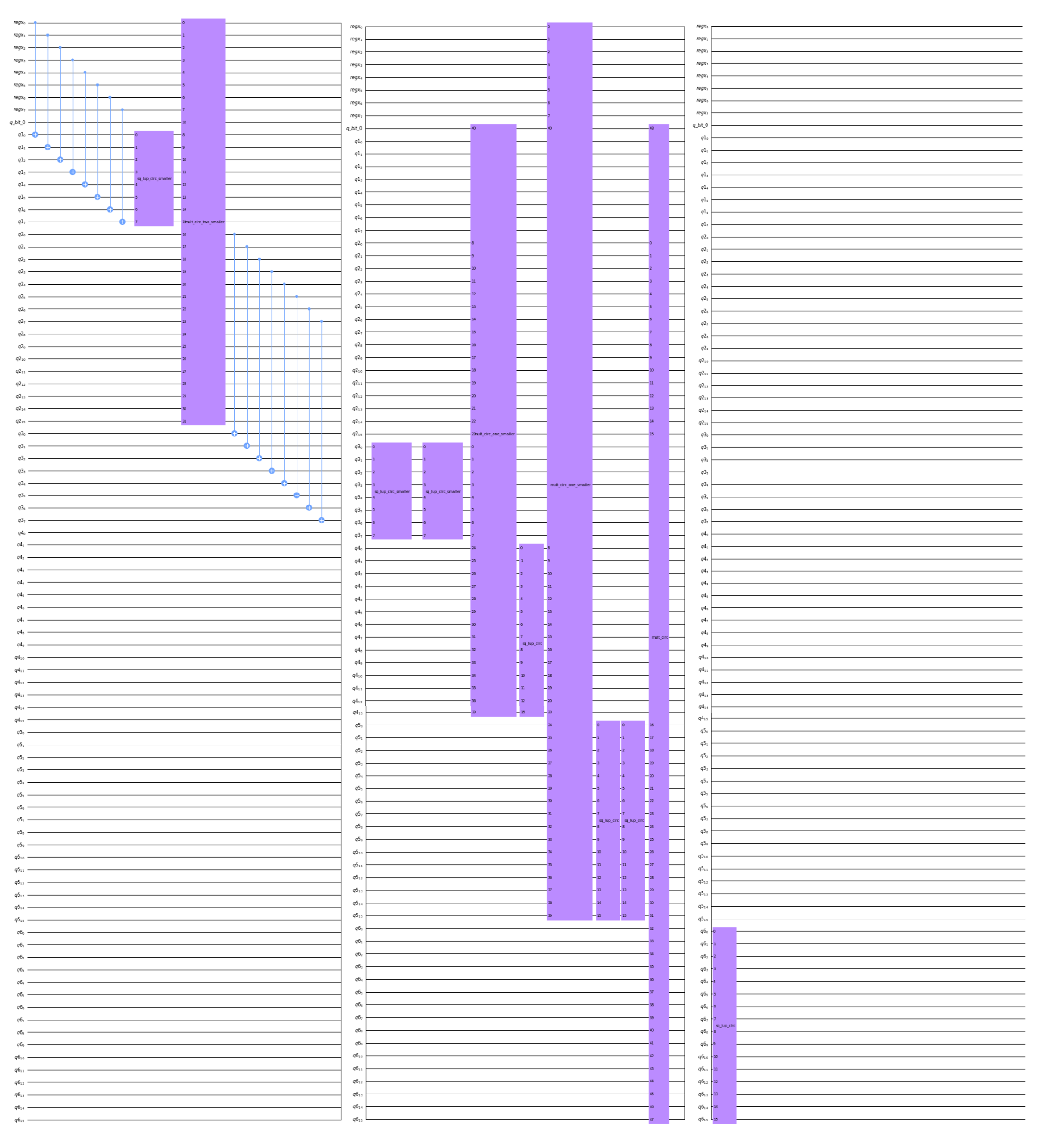}
  \caption{Our Qiskit's Circuit Construction of Our Method's FLT-based Inversion.pdf}
  \label{fig:qiskit_currentwork}
\end{figure}

Algorithm \ref{algo:flt} is translated to quantum circuit construction following the pseudocode given in the Appendix.
In the algorithm, the placement of the operations (e.g., squarings, multiplications, etc) on the quantum registers are shown. Additionally, \textit{anc} in the pseudocode refers to the list of ancillary registers, namely \textit{anc1}, \textit{anc2}, and so on. The corresponding quantum circuits for degree-8 inversions for the previous work and this work are shown in Figures \ref{fig:qiskit_prevwork} and \ref{fig:qiskit_currentwork}. For the higher bits, due to a very long circuit, the pictures are not displayed in this paper. 

\begin{figure}[h]
    \centering
    \includegraphics[width=\linewidth]{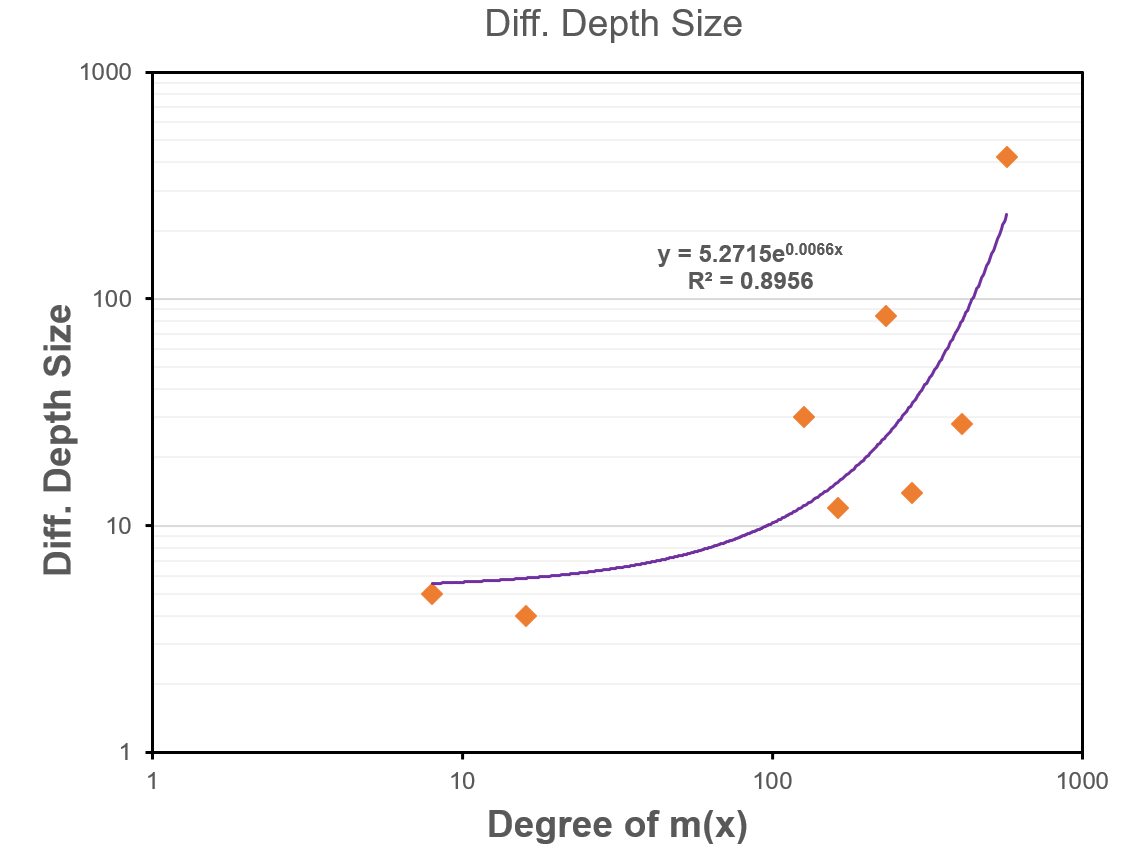}
  \caption{Depth size difference}
  \label{fig:depth_diff}
\end{figure}

\begin{figure}[h]
    \centering
    \includegraphics[width=\linewidth]{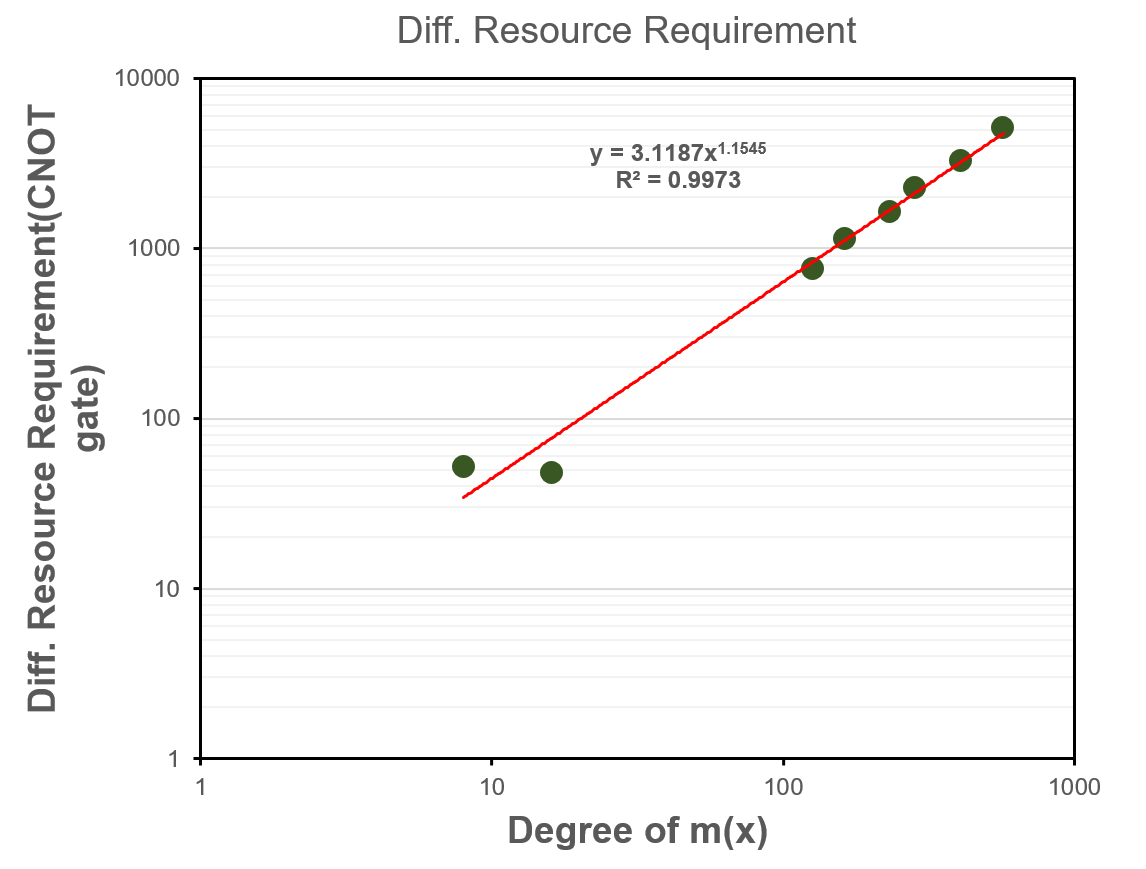}
  \caption{CNOT size difference}
  \label{fig:CNOT_diff}
\end{figure}

The result of the experiment is presented in Table \ref{tab:eval}. In all cases, our evaluations decomposed higher-level operations to CNOT, $T$, $T^\dagger$, and Hadamard gates only, all with equal gate count except CNOT gates for both the previous work and ours. As shown, the number of CNOT gates in our proposed method are lower than the previous work. This also results in a lower overall depth of the circuit. Additionally, it can be inferred that the larger the inversion bit size (which is same as the degree of $m(x)$), the larger the savings obtained from our inversion, as shown in Figures \ref{fig:depth_diff} and \ref{fig:CNOT_diff}. Note that our method comes with a tradeoff of higher circuit width (i.e., qubit size).

\begin{table*}[]
\caption{Comparison of FLT-based Inversion of the Previous Work \cite{banegas2021concrete} and This Work Based on Qiskit Simulation Result}
\centering \label{tab:eval}
\begin{tabular}{|c|cc|cc|cc|}
\hline
\multirow{2}{*}{\textbf{n}} & \multicolumn{2}{c|}{\textbf{Qubit Count}} & \multicolumn{2}{c|}{\textbf{CNOT Count}} & \multicolumn{2}{c|}{\textbf{Overall Depth}} \\ \cline{2-7} 
 & \multicolumn{1}{c|}{\textbf{Previous Work \cite{banegas2021concrete}}} & \textbf{This Work} & \multicolumn{1}{c|}{\textbf{Previous Work \cite{banegas2021concrete}}} & \textbf{This Work} & \multicolumn{1}{c|}{\textbf{Previous Work \cite{banegas2021concrete}}} & \textbf{This Work} \\ \hline
8 & \multicolumn{1}{c|}{73} & 89 & \multicolumn{1}{c|}{1856} & 1804 & \multicolumn{1}{c|}{810} & 805 \\ \hline
16 & \multicolumn{1}{c|}{209} & 257 & \multicolumn{1}{c|}{10014} & 9966 & \multicolumn{1}{c|}{2565} & 2561 \\ \hline
127 & \multicolumn{1}{c|}{2922} & 3684 & \multicolumn{1}{c|}{1074196} & 1073434 & \multicolumn{1}{c|}{31267} & 31237 \\ \hline
163 & \multicolumn{1}{c|}{3098} & 4239 & \multicolumn{1}{c|}{1484366} & 1483225 & \multicolumn{1}{c|}{53730} & 53718 \\ \hline
233 & \multicolumn{1}{c|}{4894} & 6525 & \multicolumn{1}{c|}{3306274} & 3304643 & \multicolumn{1}{c|}{58999} & 58915 \\ \hline
283 & \multicolumn{1}{c|}{6510} & 8774 & \multicolumn{1}{c|}{5431582} & 5429318 & \multicolumn{1}{c|}{161071} & 161057 \\ \hline
409 & \multicolumn{1}{c|}{9408} & 12680 & \multicolumn{1}{c|}{11148086} & 11144814 & \multicolumn{1}{c|}{111151} & 111123 \\ \hline
571 & \multicolumn{1}{c|}{15418} & 20557 & \multicolumn{1}{c|}{25778322} & 25773183 & \multicolumn{1}{c|}{200641} & 200217 \\ \hline
\end{tabular}
\end{table*}

\subsection{Discussions}
\label{sec:discussions}
This proposed work focus on minimizing the depth of the circuit as the main metric. In other words, this work wishes to achieve a time-efficient circuit. This is achieved by reducing the number of CNOT gates in the circuit by employing a complete waterfall approach to eliminate the use of the inverse squaring circuits ($SQUARE^{-1}$) used in the previous work, i.e., \cite{banegas2021concrete}. Their removal contributes to an overall lower circuit depth.

In this subsection, we note several considerations that may arise from our proposed method. Firstly, one may think that the depth reduction is not very large. However, an inversion circuit is rarely employed as a standalone circuit. Rather, it is usually incorporated to a larger circuit performing a more complex operation. For instance, one of the prominent use of quantum inversion circuit is in the Shor's ECDLP for cracking ECC \textemdash currently for quantum cryptanalysis and resource estimation, and possibly for the real cracking in the future of fault-tolerant quantum computation \textemdash specifically, as one of the components in the \textit{point addition} operation. Further, the point addition is used iteratively (repeatedly) in the \textit{scalar multiplication} operation. In the long run, more depth savings can be obtained. 

Secondly, one might question whether the overall depth, as the primary metric used in this paper, is of importance. Indeed, focusing on T depth (and T-count) reduction is very beneficial \cite{reversible_de2021reducing} due to the T-gate's criticality in the Fault-Tolerant Quantum Computation (FTQC) \cite{gheorghiu2019benchmarking}, thus has been celebrated as the focus of the recent research efforts in quantum computing and cryptanalysis. Nevertheless, as stated in \cite{reversible_de2021reducing}, the improvements in T depth and T-count are frequently accompanied by an increase in other resources, such as the number of controlled-NOT gates (CNOTs) or the overall depth of the circuit. This extra cost is not insignificant and can have an impact on a quantum algorithm's final outputs \cite{reversible_de2021reducing}. Furthermore, as discussed in \cite{maslov2016optimal}, the cost of a quantum circuit should not be reduced to the T-cost. Contrarily, it is critical to reduce secondary resources to a minimum while maintaining the T-cost unchanged \cite{reversible_de2021reducing}, like the purpose of our work. 
Another supporting argument come from a recent study on quantum cryptanalysis of Grover's algorithm on a block cipher titled KATAN, which states that NIST has not specified any limits on circuit width or the use of ancilla qubits whereas maximum depth of a circuit has been discussed \cite{rahman2022grover}. This strengthens the importance of depth, the parameter that relates to decoherence of aquantum systems and corresponds to time complexity of a quantum computer.


Another consideration that may arise is about the number of qubits. Compared to the previous work, our variant of quantum FLT-based inversion does employ larger number of qubits. However, it comes from the ancilla registers, which are also present in the FLT inversion.
On quantum computers, the ancilla registers will need to be uncomputed at the end of the computation. 
That is, the ancilla registers will be returned to state $|0\rangle$ by applying the inverse gates in reverse order.
This uncomputation results in twice the depth of the original circuit, as in \cite{parent, dutta}, whereas the ancilla qubits will be cleared up. Thus, in certain cases, reducing the depth in the large-scale future quantum computer can be more important. Additionally, going back to the discussion of our approach, the saving on depth will be doubled.  

Furthermore, looking at the progress of quantum hardware development, particularly in the most popular one, i.e., superconducting qubit, the physical qubit size has seen a significant increase in the past few years \textemdash and is quite promising to increase much more rapidly in the future \textemdash whereas progress in the coherence time (relates to depth or time complexity) is not as fast. Combined with the progress in the quantum error correction (QEC), achieving a similar increase in logical qubit may be viable in the future.

To conclude, bearing in mind that this proposed inversion circuit does not incur larger cost in both T count and T-depth, this construction can be employed as an alternative to the existing quantum inversion circuit when time complexity is the main consideration.




\section{Conclusion}
\label{sec:conclusion}
In this paper, we have presented our approach to construct FLT-based quantum inversion circuit with lower depth for use in binary elliptic curves. We offer an alternative variant from the prior work \cite{banegas2021concrete} to perform inversion, a crucial subcircuit of division circuit and the most expensive sub-part of computing the scalar multiplication in the quantum binary elliptic curve. 
In detail, the proposed design employs different approaches at the first and second stages of Banegas et al., \cite{banegas2021concrete}'s elaboration of Itoh-Tsujii’s FLT inversion algorithm. We utilize a different arrangement of the quantum circuit and move the uncomputation of squarings to the end to eliminate the intermediate uncomputation, minimizing the number of CNOT gates in the circuit. As a result, a lower overall gate count and circuit depth can be achieved. For verification, we implemented both our method and the previous work's FLT approach in IBM Qiskit quantum computer simulator and compared the resource requirement for B-233, B-283, B-409, and B-571 as the standard for binary elliptic curve, which confirms that a lower overall gate count and circuit depth can be achieved when our approach is used in the perspective of quantum cryptanalysis of ECC. Additionally, our approach maintains the same T-depth as the previous work, which is of additional advantage. Note that our approach does incur a higher number of auxiliary qubits. Nevertheless, the fact that uncomputation must also be performed in any case, meaning that the auxiliary qubits will still need to be uncomputed at the end, gives our approach a leverage when one wants to pursue a lower depth implementation.

\ifCLASSOPTIONcaptionsoff
  \newpage
\fi


\printbibliography
\onecolumn
\section*{Appendix: Python-style Pseudocode for Circuit Construction in Qiskit }
\begin{figure*}[htb] \label{algo:flt_pseudocode}
    \centering
    \includegraphics[width=0.83\linewidth]{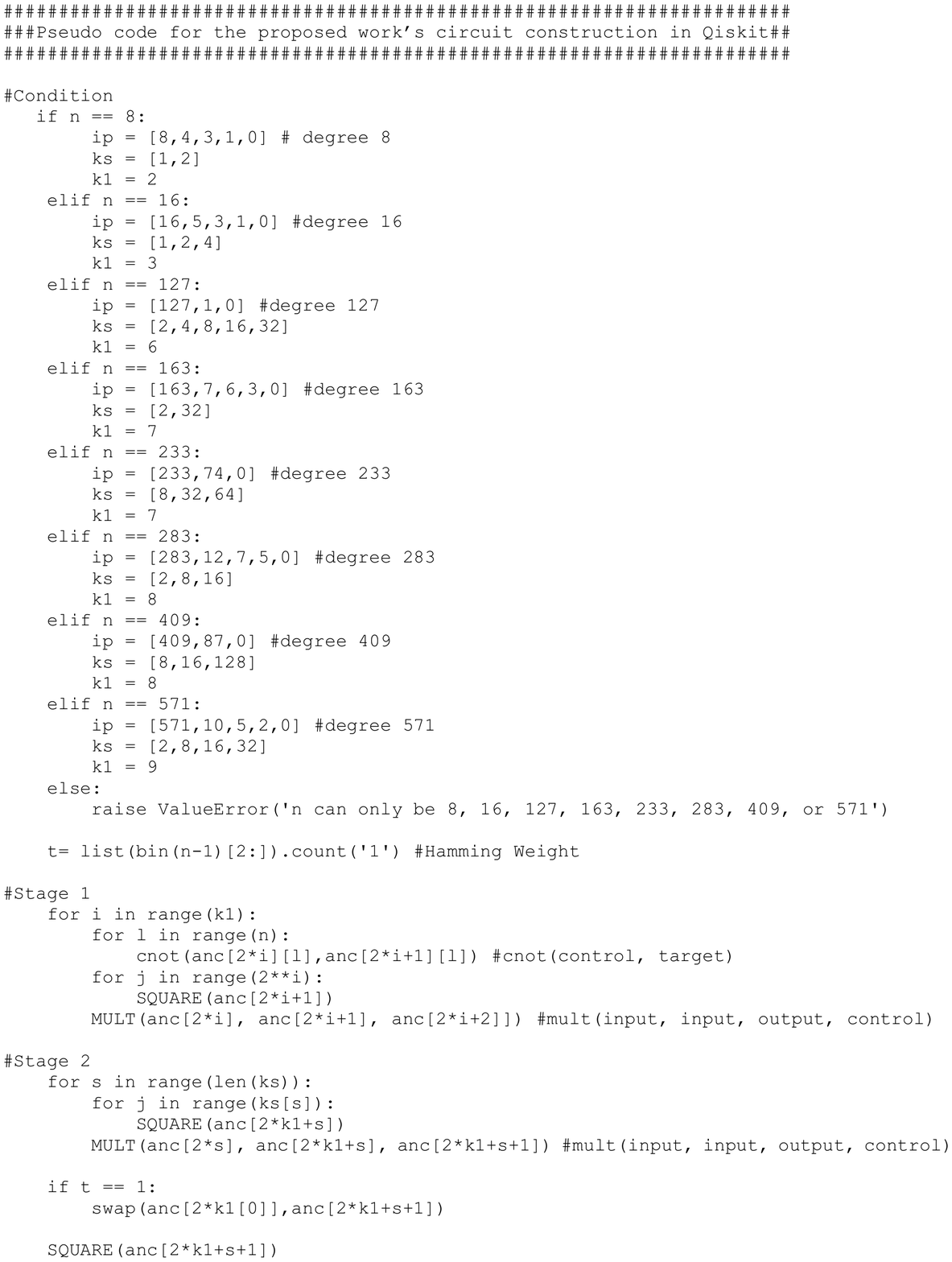}
\end{figure*}
\end{document}